# Child Care Provider Survival Analysis

*Phillip Sherlock[1], Herman T. Knopf[2], Robert Chapman[1], Maya Schreiber[2], Courtney K. Blackwell[1]*

[1] Northwestern University, [2] University of Florida

**Abstract**

The aggregate ability of child care providers to meet local demand for child care is linked to employment rates in many sectors of the economy. Amid growing concern regarding child care provider sustainability due to the COVID-19 pandemic, state and local governments have received large amounts of new funding to better support provider stability. In response to this new funding aimed at bolstering the child care market in Florida, this study was devised as an exploratory investigation into features of child care providers that lead to business longevity. In this study we used optimal survival trees, a machine learning technique designed to better understand which providers are expected to remain operational for longer periods of time, supporting stabilization of the child care market. This tree-based survival analysis detects and describes complex interactions between provider characteristics that lead to differences in expected business survival rates. Results show that small providers who are religiously affiliated, and all providers who are serving children in Florida's universal Prekindergarten program and/or children using child care subsidy, are likely to have the longest expected survival rates.

Keywords: child care, survival analysis, resource allocation, educational economics

## 1. Introduction

The aggregate ability of child care providers to meet local demand for child care is linked to employment rates in many sectors of the economy. A national panel survey revealed that nearly 20% of working parents had their work schedules disrupted due to difficulties maintaining child care during the COVID-19 pandemic, and only 30% of all working parents had a backup-care arrangement (Modestino et al., 2021). Amid growing concern regarding provider sustainability due to the COVID-19 pandemic, state and local governments have received large amounts of new funding to better support provider stability (CARES, 2020; CRRSA, 2020; ARPA, 2021). These investments have provided all states with unprecedented amounts of funding to be disbursed to child care providers at the discretion of the states, within specified guidelines.

In particular, the child care stabilization grants in the American Rescue Plan Act (ARPA; 2021) are intended to increase provider longevity to both maintain and improve child care infrastructure. These require states to provide funding directly to child care providers; however, states are given discretion in how they choose to allocate this funding across providers (Office of Child Care, 2021). In order to ensure optimal allocation of these funds, it is important to prioritize investment in providers who are expected to remain operational over many years.

In Florida, there are two primary state agencies who are tasked with monitoring and engaging with child care providers: (1) the Department of Children and Families (DCF); and (2) the Florida Division of Early Learning (DEL) (previously the Office of Early Learning; OEL). DCF is responsible for monitoring child care provider licensing as well as basic health and safety standards in most counties (Florida Department of Children and Families, n.d.). DEL is responsible for administering the Child Care Development Fund (CCDF) funded child care subsidy program and Florida's Voluntary Prekindergarten (VPK) program, which is operated in private and public sector settings (Florida Division of Early Learning, n.d.a.).

Given the complex, ever-evolving landscape of the early care and education market, it should be noted that this study focuses on providers who interact with any of the following three programs: (1) Florida School Readiness (SR); (2) Florida Voluntary Pre-Kindergarten (VPK); and (3) Florida Gold Seal Quality Care (GS). Florida's SR program is the state's CCDF subsidized child care program, which supports low-income families in accessing child care (Florida Division of Early Learning, n.d.b.). In order to be eligible to serve subsidy recipients in the state, providers must have an SR contract with DEL (Florida Division of Early Learning, 2022a). The VPK program provides universal pre-kindergarten to all families in Florida (Florida Division of Early Learning, n.d.c.). In order to provide VPK services a provider must be contracted with DEL and meet basic requirements to administer services (Florida Division of Early Learning, 2022b). The Florida GS program provides financial incentives to providers who have obtained accreditation from approved accrediting agencies and are subsequently designated as "Gold Seal" providers (Florida Division of Early Learning, n.d.).

In this study, we performed an exploratory survival analysis using administrative child care provider records from Florida, which include providers who served children through a variety of public and private payment mechanisms. The purpose of this exploratory study is to investigate characteristics of Florida child care providers that lead to differential business survival rates. The results of this analysis will offer an initial outlook on the salient features of child care providers that are associated with increased survival rates. This is a necessary first step in understanding historical business life cycles among providers, taking into account features such as type of provider (e.g., home-based, center-based), capacity (i.e., number of children a provider can legally serve), and whether these providers participate in state-funded initiatives. The main aim of this study was to support state decision makers as they implement new policies and allocate funding to support durable child care infrastructure.

## 2. Materials and Methods

### *2.1 Data*

Two data samples were retrieved from the Florida DCF website, the state's child care licensing agency. The first sample, referred to as the "open provider dataset," included a list of all licensed providers who were operating at the time of data collection (November 2021) with 13,102 records. The second sample, referred to as the "closed provider dataset," included all licensed providers that were closed at the time of data collection (November 2021) with 5,941 records. In both datasets, providers were organized and identified by a unique DCF provider identification number (DCF ID) which was used to merge the two files, creating a unified list of open and closed providers. Table 1 describes all of the predictor variables that were considered during model development.

Table 1. Covariates Originally Present in Open and Closed Providers Datasets

| Type | Variable Name | Measurement Scale | Description | Categories/Range |
|---|---|---|---|---|
| Program Type | | | | |
| | Program Type | Categorical, Nominal | Program type of child care provider, only one listing possible | Child Care Facility, Family Day Care Home, Large Family Child Care Home, Informal |
| | Faith Based | Categorical, Dichotomous | A positive/negative endorsement of whether a child care provider is faith based or not | Yes, No |
| | Urban Zoned | Categorical, Dichotomous | A positive/negative endorsement of whether a child care provider is urban zoned or not | Yes, No |
| | School Aged Only | Categorical, Dichotomous | A positive/negative endorsement of whether a child care provider enrolls only school-aged children (children ages 6-13) | Yes, No |
| Provider Size | | | | |
| | Capacity | Interval | The upper limit of how many children a child care provider can support, based on Florida Department of Children and Families square footage guidelines | 0-999 children |
| Provider Status | | | | |
| | License Status | Categorical, Nominal | A list of various license statuses, including valid and invalid license statuses | Exempt, Illegal, Licensed, Registered, Substantial Compliance |
| Provider Quality | | | | |
| | Gold Seal Status | Categorical, Nominal | The status of participation in Florida's Gold Seal program which provides a 'Gold Seal' quality designation to providers with an accreditation from specified accreditation agencies. | Active, inactive, terminated |
| Provider Subsidy | | | | |
| | School Readiness Status | Categorical | Child care provider status in the Florida School Readiness (SR) child care subsidy program | Active, applied, terminated |
| | Voluntary Pre-Kindergarten | Categorical, Dichotomous | Whether or not a child care provider participates in the Voluntary Pre-Kindergarten (VPK) program | Yes, No |

*2.2 Deduplication Process*

Data were reviewed for duplicates within each sample. One duplicate closure record was removed. Providers' DCF IDs were then reviewed between the two data files to confirm the absence of overlapping records (i.e., providers could be either currently open or currently closed, but not both). Five overlapping records (matched by DCF IDs) were removed.

*2.3 Delimitation*

The closed dataset included providers with origination years from 1957 to 2021 and closure years from 2017 to 2021. The open providers dataset included providers with origination years from 1961 to 2021 and no closure years. Although the open and closed datasets include a broad range of years, we chose to only include data from the past ten years, 2012 to 2021, for two reasons. First, the quantity of provider records and availability of covariate data greatly improved in the past ten years. Second, in Florida, 19.7%, 50.6%, and 65.4% of businesses fail within the one, five, and ten years, respectively (U.S. Bureau of Labor Statistics, n.d.). Across all states, there is a tapering effect in the rate of business failure, such that for each additional year after five years, the probability of failure grows at a much slower rate than in the first five years (U.S. Bureau of Labor Statistics, n.d.). Due to the data delimitation used for this exploratory study, the longest possible business survival for a child care provider is 10 years.

Additionally, child care providers who are also public schools were excluded from the analyses, as the circumstances leading to public school closures were assumed to be different than those leading to failed child care provider businesses.

*2.4 Covariate Selection and Calculation*

In total, 11 predictors were included in the survival tree analysis. Provider origination year, program type, license status, GS, SR, VPK, Head Start status, and provider capacity were included as predictors in the model. Variables indicating whether a provider is faith based, urban zoned, and only enrolls school-aged children (older than 5 years of age) were also included in the model.

For all records, the 'years of operation' variable was calculated using the provider origination year and closure year. This variable ranged from 0 to 10 years of operation, as the earliest provider origination year included in the dataset was from 2012 and the last data collection point was 2021. Table 2 describes the data from open and closed child care providers that were used to fit the Optimal Survival Tree model. The majority of providers were licensed (n=4,454, 80.40%) child care facilities (n=3,859, 69.66%) not participating in the VPK (n=4,233, 76.41%) or Head Start programs (n=5,330, 96.21%) with an average capacity of 65 children (mean capacity = 64.7).

Table 2. Descriptive Statistics for Child Care Providers used in Optimal Survival Tree Models (n=5,540)

| License Status | n | % | Voluntary Pre-Kindergarten | n | % |
|---|---|---|---|---|---|
| Exempt | 576 | 10.40% | Yes | 1307 | 23.59% |
| Licensed | 4454 | 80.40% | No | 4233 | 76.41% |
| Registered | 499 | 9.01% | | | |
| Substantial Compliance | 11 | 0.20% | Faith Based | n | % |
| | | | Yes | 500 | 9.03% |
| Program Type | n | % | No | 5040 | 90.97% |
| Child Care Facility | 3859 | 69.66% | | | |
| Family Day Care Home | 1462 | 26.39% | Head Start | n | % |
| Large Family Day Care Home | 219 | 3.95% | Yes | 210 | 3.79% |
| | | | No | 5330 | 96.21% |
| Gold Seal Status | n | % | | | |
| Active | 431 | 7.78% | Urban Zoned | n | % |
| Inactive | 133 | 2.40% | Yes | 19 | 0.34% |

| | | | | | | |
|---|---|---|---|---|---|---|
| No | 4975 | 89.80% | | No | 5521 | 99.66% |
| Terminated | 1 | 0.02% | | | | |
| | | | | School Aged Only | n | % |
| School Readiness Status | n | % | | Yes | 259 | 4.68% |
| Active | 2024 | 36.53% | | No | 5281 | 95.32% |
| Applied | 28 | 0.51% | | | | |
| No | 2207 | 39.84% | | Origination Year | *median, range* | |
| Terminated | 1281 | 23.12% | | | 2016, 2012-2021 | |
| | | | | Capacity | *mean(sd), range* | |
| | | | | | 64.7(70.7), 0-758 | |

## *2.5 Optimal Survival Trees*

Tree-based models, such as the Classification and Regression Tree algorithm (CART; reiman et al., 1984) have become popular due to their ability to model high-dimensional, complex interactions and for their ease of interpretability. This approach simultaneously identifies the most meaningful predictors of an outcome and their respective threshold values that divide the sample into terminal nodes (i.e., subgroups formed following each sequence of splits). However, survival analyses necessitate the handling of missing observations that arise in the context of censored data, which requires a specific type of approach that is not achievable with traditional tree-based regression models. Instead, we used the Optimal Survival Tree (OST) algorithm, an approach developed by Bertsimas and Dunn (2017), which uses modern mixed-integer optimization. CART first splits the data based on the predictor that leads to the highest reduction in prediction error and is therefore susceptible to local minima. The tree resulting from the CART procedure may not represent the optimal structure for prediction. Additionally, the CART approach does not allow for the modification of splits once they enter the model. OSTs, however, are constructed in a single step, wherein each split is determined with full knowledge of all other splits. Furthermore, the OST algorithm produces a single, interpretable tree, which in many different applications have been shown to perform similarly to cutting-edge, highly predictive methods like random forests (Breiman, 2001) and gradient boosted trees (Friedman 2001; 2002) that do not have the same interpretability.

The specific arguments and options used in the Interpretable AI Optimal Survival Tree modeling were based on the recommended settings and provided examples from Interpretable AI. A training/testing split of 75%/25% was used for model building and validation (i.e., out-of-bag [OOB] prediction). We made one a priori change to the default OST settings, wherein we increased the default minimum terminal node size from 1 to 75 (which represents 1% of the analytic sample) in order to increase the likelihood of robust findings and decrease the likelihood of spurious subgroups that would compromise OOB prediction.

The "max depth" parameter was tuned in the range of four to ten with the complexity parameter automatically tuned. Max depth refers to the maximum number of splits between the root node and any terminal node. It is often the case that the depth of tree-based models can yield complex interactions that go beyond the capabilities of regression-based modeling with respect to the interpretability of high-dimensional, multi-way interactions (e.g., five-, six-, and seven-way interactions). However, as it was our intention to yield an interpretable tree with optimal predictive ability, we balanced prediction error in the tuning process with interpretability and sought a max depth that was sufficiently predictive, yet manageable with respect to interpretation and generalizability. Harrell's C-index was used to assess model fit. Harrell's C, an adaptation of the concordance statistic from logistic regression, was used as a goodness-of-fit measure in our final survival model (Harrell et al., 1982). Harrell's C, in the case of survival analysis, and specifically this study, refers to the probability of our model assigning a higher survival probability to the provider that actually stayed in business longer. The C-index which ranges from 0 to 1 can also be thought of in terms of a coin flip. A C value equal to 0.5 means that the model correctly identifies the correct survivor, between all pairs, 50% of the time. It follows that a value of 1 would mean perfect prediction between all allowable comparisons (i.e., pairs of providers for whom we know which provider closed first).

The random nature of splitting the dataset into test/train subsamples and starting locations for parameter estimation used in optimal survival trees means that a "random seed" must be set to precisely replicate the model results. To detect potentially spurious results, once the final model parameters are identified, the random seed may be reset (i.e., set to another value), allowing for the evaluation of a second model with the same input parameters. The two models with different random seeds can be evaluated to determine if the same relationships between variables re-emerge with new starting values and a different random split between test and train data subsets.

## 3. Results

### *3.1 Model Selection and Fit*

The max-depth tuning procedure, with automatically tuned complexity parameter, yielded a set of seven best-fitting survival tree models based on a maximum possible number of nodes from four to ten. Table 3 shows Harrel's C values for the training and testing data subsets of each best-fitting model, organized by the maximum number of nodes. The OST algorithm selected the same model of best fit when the maximum numbers of nodes were set to 7 and 8, as well as 9 and 10, as indicated by identical Harrell's C values and complexity parameters. These values are shown in Table 3. The automatically tuned complexity parameters were very small across all models with alpha values ranging from 0.00000158 to 0.000905. Based on a comparison of the Harrell's C values across the models, we generally found values of Harrell's C in the range of .85 to .90, with all the C values for the testing samples greater than their associated training-sample values. Ultimately, we chose the model with a max depth equal to five because the greatest increase in Harrell's C occurred from a max depth of four to five—the added predictive value decreased with each additional max-depth integer, as indicated by a gray vertical bar in Figure 1. Furthermore, this drop-off with respect to the increase in the predictive abilities of the more complex models was an indication that a simpler, more easily interpretable tree with a max depth equal to five was the preferred model. A final max depth equal to five was selected.

### *3.2 Model Reseeding*

To ensure that the results were reproducible and consistent, the max depth model was "reseeded", meaning that the random seed set to calculate the models (seed = 1) was reset to different, random value (seed = 352). The original model and the reseeded model yielded no substantive differences compared to the original model (e.g., the same model variables appeared in both the original and reseeded models). In the final tree with max depth equal to five, Harrell's C for the training data was 0.87 and the test data was 0.87, with an automatically tuned complexity parameter (alpha) of 0.0002. Related to Harrell's C, this means that in 87% of the compared pairs of providers, the model correctly assigned lower expected survivals to the providers who indeed were the first to close among all the allowable pairs. The reseeded model was selected as the final model and is explained below.

Table 3. Output Parameters for Candidate Optimal Survival Tree Models

| Max. # of Nodes in Modeling | 4 | 5 | 6 | 7 | 8 | 9 | 10 |
|---|---|---|---|---|---|---|---|
| Complexity Parameter (Alpha) | 0.000844 | 0.000905 | 0.000293 | 0.00000158 | 0.00000158 | 0.000172 | 0.000172 |
| Harrell's C | | | | | | | |
| Train | 0.849 | 0.867 | 0.872 | 0.875 | 0.875 | 0.880 | 0.880 |
| Test | 0.863 | 0.880 | 0.883 | 0.884 | 0.884 | 0.891 | 0.891 |

Figure 1. Harrell's Concordance Statistic for Models by Max Number of Nodes

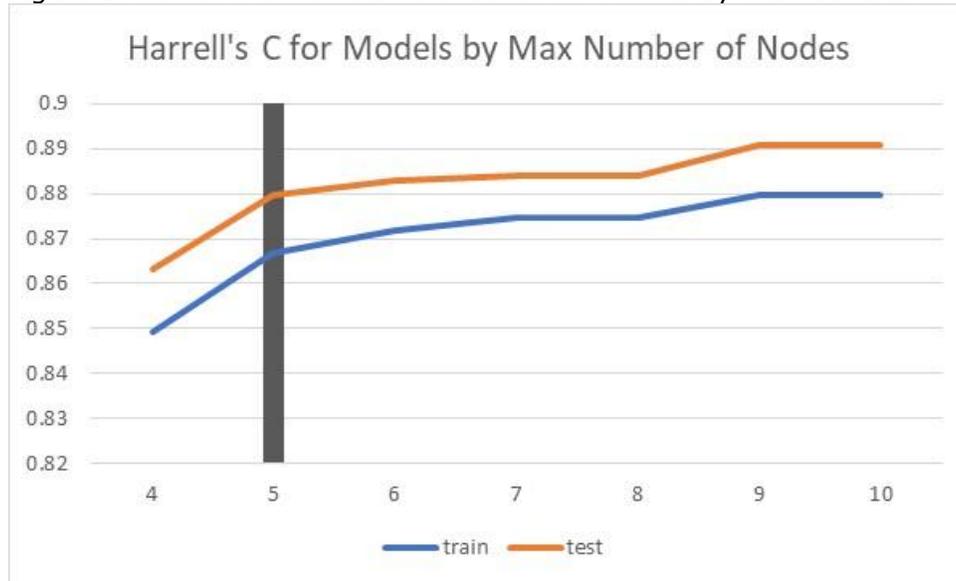

Figure 2. Final Reseeded Model with Max Number of Nodes Equal to Five

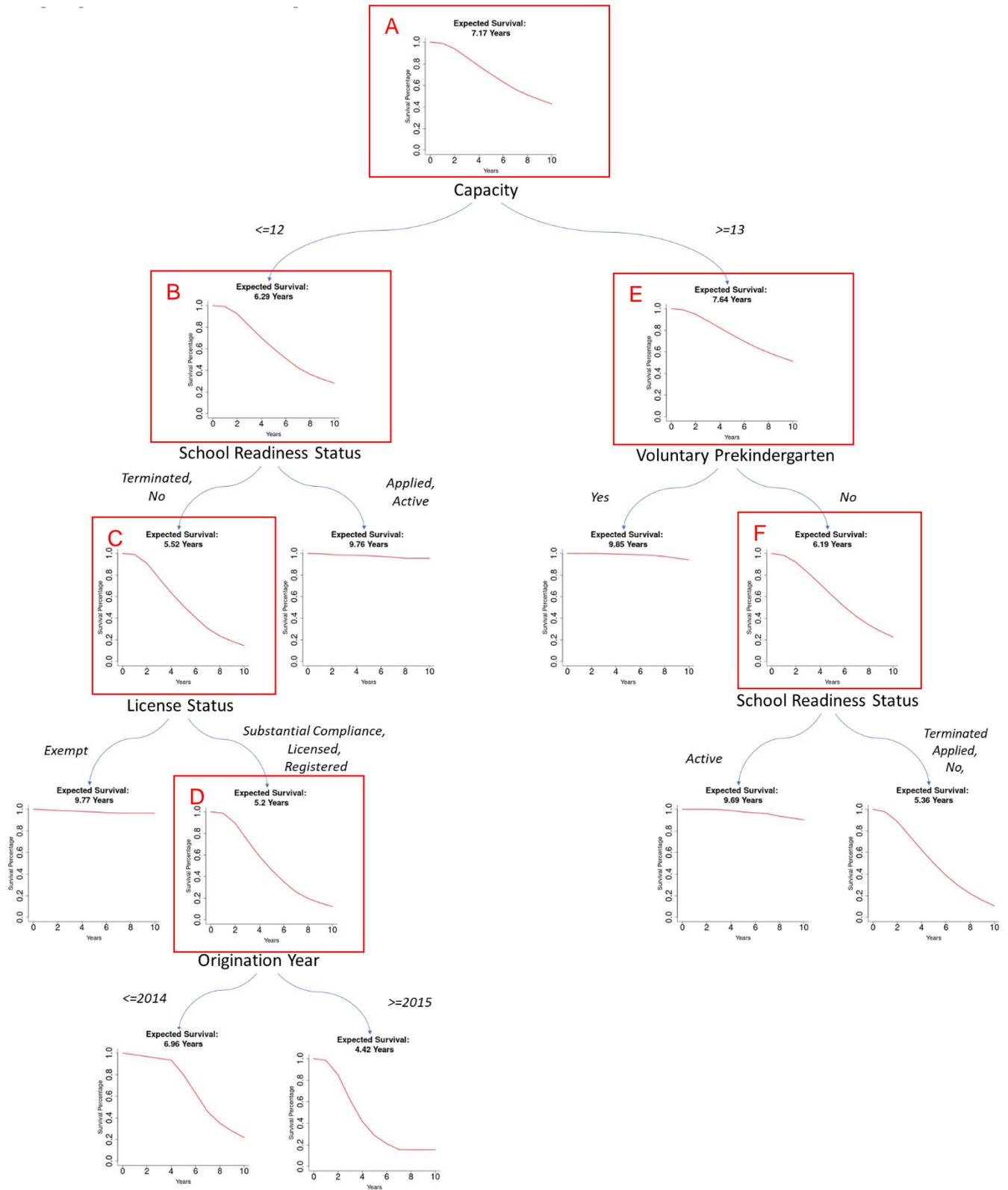

9.8 years. The remaining providers, those with a license status of registered, licensed, or substantial compliance (Figure 2, Node D), were further split by origination year. Providers who originated in or prior to 2014, were not a part of the SR program, and were not license exempt, have an expected survival of 7 years. Providers who have these same characteristics, with the exception of having an origination date after 2014, have the lowest expected survival across all providers (4.4 years).

*Larger Providers (Capacity ≥ 13)*

As a group, larger providers have an expected survival of 7.6 years (Figure 2, Node E). Among these providers, those who are active in the VPK program have a greater expected survival of 9.8 years compared to larger providers who are not active in the VPK program (6.2 years, Figure 2, Node E). Larger providers who are not active in the VPK program but are active in the SR program have an expected survival of 9.7 years. Providers who are not active in either program have an expected survival of 5.4 years.

**Discussion**

The purpose of this study was to explore how expected child care business survival varied among providers in Florida between 2012 and 2021. Results from the survival analysis uncovered differential vulnerabilities and protective factors among providers with respect to business longevity. Small providers who were active in or who applied for the SR program and/or had an exempt license status have an expected survival of 9.8 years. Conversely, exempt, small providers who were not eligible to serve SR recipients or had been terminated from SR had an expected survival of 5.2 years. Among larger providers, those who were active in the SR program and/or active in the VPK program, had the highest expected survival, which was at least 9.7 years. Conversely, larger providers who did not contract with VPK or SR, had an expected survival of 5.36 years. Taken together these results highlight the importance of participation in state-funded initiatives across both center-based and home-based providers. The results also point to potential protective factors related to license exemption among small providers (i.e., capacity < 13) who were either not participating in SR or who were terminated from the program. It should be noted that in this sample all license-exempt providers were exempt due to religious affiliations. One possibility is that faith-based providers may have stronger ties to local communities given they are associated with religious institutions within their respective communities. Another possibility is that the operational costs of faith-based providers are subsidized by their affiliated organization (e.g., rent, personnel, insurance, tax exemption).

In regard to the main aim of this study, which was to provide guidance to support Florida's decision makers in allocating funding to support durable child care infrastructure, results highlight the importance of interventions aimed specifically at increasing provider eligibility to serve children funded through state-directed programs. As a matter of business practice, providers who contracted with state agencies tended to have greater expected longevity, and this deserves further examination. One possibility is that providers who are able to contract with the state and participate in the SR and/or VPK program(s) have demonstrated the organizational ability and business acumen that are becoming of business sustainability. Another possibility is that a provider can maximize the likelihood of maintaining full service capacity by expanding their customer base to include families who receive subsidized care. Theoretically, even if only a small percentage of child care slots are utilized by subsidy recipients in many providers, it would seem to be in the best interest of providers to establish and maintain state contracts in order to maintain the greatest utilization of capacity, which in turn would lead to increased, sustained revenue. Therefore, the results of this study, as well as logic, dictate that it is in the best interest of providers to contract with state agencies that issue subsidized child care vouchers, in order to access the entire population of child care users, particularly the ones most in need of the service.

Interestingly, while Gold Seal status was included in the analysis as a predictor, it was not identified as a salient provider characteristic that predicted business longevity. This should not be taken to mean that provider quality is not important. Rather, this finding should be strictly interpreted in the context of the survival analysis—after accounting for other salient provider features (e.g., SR participation, VPK participation), provider participation in GS does not seem to be associated with differences in business longevity. Furthermore, GS participation and more broadly, provider quality, is certainly an important topic, but seems to be differentiable from business sustainability.

## 4. Limitations and Future Directions

Despite having uncovered salient features that are associated with child care provider longevity, the survival analysis did not include information related to the financial perspectives (e.g., earnings reports, non-subsidized and subsidized enrollment) of the included providers, simply because this information was not available. That said, the results of this study would be further complemented by provider financial information that can be used to generate and test more specific hypotheses about the causes of child care provider closures and differential longevity.

While a primary aim of this study was to inform Florida early child care policymakers, findings may have broader applicability to other states. Related to state-funded child care funding, this study points to the potential buffering effect of being contracted to serve children funded by state-directed programs. Related to license-exempt providers (i.e., faith-based organizations), this buffering effect may also be present in providers operating in other states with similar license-exemption models. Ultimately, the variability in the structure and implementation of child care systems across states necessitates state-level investigation of the salient characteristics of provider longevity to help inform decision-making and resource allocation.

## 5. Acknowledgements

We would like to express our sincerest gratitude to our partners from the Florida Department of Children and Families, Office of Licensing, and the Florida Department of Education, Division of Early Learning. This research is supported by the Division of Early Learning through the Preschool Development Grant Birth through Five Initiative (PDG B-5) Number 90TP0004-02-00 from the Office of Child Care, Administration for Children and Families, U.S. Department of Health and Human Services. Its contents are solely the responsibility of the authors and do not necessarily represent the official view of the United States Department of Health and Human Services, Administration for Children and Families.

## 6. Declaration of Interest Statement

We have no interests to declare.